\pdfoutput=1

\documentclass[aps,prb,reprint,showpacs,superscriptaddress,floatfix]{revtex4-2}
\usepackage{graphicx}
\usepackage{graphics}
\usepackage{xcolor}
\usepackage{amsmath}
\usepackage{tabularx}
\usepackage{amssymb}
\usepackage{comment}
\usepackage{amsfonts}
\usepackage{xcolor}
\usepackage{dcolumn}
\usepackage{dsfont}
\usepackage{latexsym}
\usepackage{rotating}
\usepackage{color}
\usepackage{latexsym}
\usepackage{bbm}
\usepackage{subfigure}
\usepackage{float}
\usepackage{epsfig}
\usepackage{epsf}
\usepackage{psfrag}
\usepackage{bm}
\usepackage{amsthm}
\usepackage{eucal}
\usepackage{mathrsfs}
\usepackage{url}
\usepackage{braket}
\usepackage{array}
\usepackage[utf8]{inputenc}
\usepackage{microtype}
\usepackage{multirow}
\usepackage{comment}
\usepackage{hyperref}
\hypersetup{
colorlinks=true,final=true,
        linkcolor=blue,
        citecolor=blue,
        filecolor=blue,
        urlcolor=blue,
}

\usepackage{xparse}
\usepackage{xspace}

\NewDocumentCommand{\Cmmm}{}
{%
  \textit{Cmmm}\xspace%
}

\NewDocumentCommand{\Imma}{}
{%
  \textit{Imma}\xspace%
}

\NewDocumentCommand{\Fmmm}{}
{%
  \textit{Fmmm}\xspace%
}

\usepackage{ulem}

\date{\today}

\begin{document}

\title{Structural stability, electronic structure, and magnetic properties of the single-layer trilayer {L}a$_{3}${N}i$_{2}${O}$_{7}$ polymorph}
\author{Shekhar Sharma}
\altaffiliation {sshar246@asu.edu}
\affiliation{Department of Physics, Arizona State University, Tempe, AZ 85287, USA}
\author{Yi-Feng Zhao}
\altaffiliation {yzhao421@asu.edu}
\affiliation{Department of Physics, Arizona State University, Tempe, AZ 85287, USA}
\author{Antia S. Botana}
\affiliation{Department of Physics, Arizona State University, Tempe, AZ 85287, USA}

\begin{abstract}

A polymorph of the bilayer nickelate La$_{3}$Ni$_{2}$O$_{7}$ that displays an alternating single-layer and trilayer (1313) stacking pattern has recently been discovered. Signatures of superconductivity under pressure have been found in this novel phase. At ambient pressure,  La$_{3}$Ni$_{2}$O$_{7}$-1313 has been reported to crystallize in three different space group symmetries  \Cmmm, \Imma, and \Fmmm. Unlike the commonly observed tilted NiO$_{6}$ octahedra in perovskite nickelates, the \Cmmm phase exhibits no NiO$_{6}$ tilts, implying that this structural feature alone may be insufficient to give rise to superconductivity in Ruddlesden-Popper nickelates. Here, we employ first-principles calculations and group theory analysis to study the pressure dependence of the structural instabilities in this single-layer trilayer La$_{3}$Ni$_{2}$O$_{7}$ polymorph. At ambient pressure, we identify multiple unstable phonon branches in the highest symmetry (\Cmmm) structure at various high-symmetry points of the Brillouin zone. Distortions associated with these instabilities lead to one of the other experimentally reported space groups (\Imma), that does display octahedral tilts. The magnetic tendencies indicate that the electronic structure of La$_{3}$Ni$_{2}$O$_{7}$-1313 at ambient pressure is dominated by the trilayer block, as the single-layer is in a Mott-insulating regime. Under pressure, a tetragonal $P4/mmm$ structure becomes stable, in agreement with experiments.

\end{abstract}

\maketitle

\section{Introduction}

Since the discovery of the  cuprates \cite{Bednorz1986}, there have been several attempts at identifying new high-T$_c$ materials with cuprate-like characteristics \cite{Norman_2016}. Given the proximity of nickel to copper in the periodic table, nickelates have been considered prime candidates in this context \cite{anisimov1999,Ni1+isnotCu2+,Chaloupka_PRL_Orbital_Order_2008,Held2009_prl,Han2011_PRL}. By drawing analogies with the cuprates in terms of their crystal structure and $3d$ electron count, superconductivity in nickelates was first discovered in epitaxial thin films of the reduced square-planar phases R$_{n+1}$Ni$_{n}$O$_{2n+2}$ (R= rare-earth), characterized by a cuprate-like sequence of $n$-NiO$_{2}$ layers separated by fluorite RO$_{2}$ blocks along the $c$-axis \cite{ZHOU_2022_exp_progress_nickelates, berit_review, mitchell2021nickelate, Gu_science_nickelates_review}. These phases have a Ni oxidation state close to 1+, isovalent to Cu$^{2+}$, and they display an electronic structure characterized by the dominant role of $d_{x^2-y^2}$ bands, in a cuprate-like fashion \cite{botana2020, hepting2020,kitatani_2023_cuprates_via_nickelates,jiang2019, review_es_layered, PRB_harrison_elecrtonic_structure_higher_order_2021}. Within this family, both the infinite-layer ($n=\infty$) compounds RNiO$_2$ (R= La, Pr, Nd)~\cite{Li2019,Osada2020_nanolett,Osada2020_prm,Li2020dome} and the quintuple-layer ($n=5$) Nd$_6$Ni$_5$O$_{12}$ material \cite{pan2021super} have been shown to be superconducting with similar T$_c$ $\sim$ 15 K.

A significant breakthrough in the field occurred in 2023
with the observation of superconductivity under pressure in the parent Ruddlesden–Popper (RP) R$_{n+1}$Ni$_{n}$O$_{3n+1}$ phases~ \cite{sun2023signatures,wang2024bulk, hou2023emergence,La4Ni3O10_anomaly_resistivity,li2023signature}, characterized by a sequence of $n$-NiO$_{6}$ perovskite layers separated by rocksalt RO blocks along the $c$-axis \cite{greenblatt1997}. The first superconducting signatures were reported in the bilayer ($n=2$) RP nickelate
La$_3$Ni$_2$O$_7$~\cite{sun2023signatures, hou2023emergence}. This material has been shown to exhibit a T$_c$ $\sim$ 80 K within a pressure range of $\sim$ 14.0 to
40 GPa~\cite{sun2023signatures, hou2023emergence}. 
Concomitant with the emergence of superconductivity,  a structural transition from orthorhombic \textit{Amam} at ambient pressure to \textit{Fmmm} (or tetragonal \textit{I4/mmm}) under pressure has been reported~\cite{sun2023signatures, wang2024structure}. This transition is associated with a change in the out-of-plane Ni-O-Ni bond angle from $\sim$ 168.0$^{\circ}$ to 180.0$^{\circ}$. 
The electronic structure of this material is characterized by the active role of bands of $d_{x^2-y^2}$ and $d_{z^2}$ character. Many studies have suggested that the straightening of the Ni-O-Ni bond angle along the $c$-axis with pressure is related to the emergence of superconductivity due to the associated enhancement in $d_{z^2}$ inter-layer coupling via the apical oxygens~\cite{luo2023bilayer, Zhang2023, Yang2023, Yang2024, sakakibara2024, christiansson2023correlated,  lechermann2023electronic, Yang2024, Zhang2024save, zhang2024structural, Lu2024,Yang2024-orbital_selective}.

Recently, a new polymorph of La$_3$Ni$_2$O$_7$  was reported ~\cite{1313Polymorphism202JACSmitchel,Puphal_PhysRevLett.133.146002,Wang_2024_Cmmm,arxivheptingmono-tri}. In contrast to the typical uniform bilayer (2222) stacking of perovskite blocks of La$_3$Ni$_2$O$_7$, the structure of this polymorph consists of a novel sequence of alternating single-layer and trilayer blocks forming a 1313 configuration. Superconductivity in this new La$_3$Ni$_2$O$_7$-1313 polymorph has also been reported under pressure with a  T$_c$ as high as that of La$_3$Ni$_2$O$_7$-2222~\cite{arxivheptingmono-tri,Puphal_PhysRevLett.133.146002,1313_anomaly}. An anomaly in the resistivity of La$_{3}$Ni$_{2}$O$_{7}$-1313 occurs at $\sim$ 180 K~\cite{1313_anomaly, 1313Polymorphism202JACSmitchel}. A kink has also been identified at a similar temperature in the magnetic susceptibility \cite{1313_anomaly}. This anomaly is hence similar to that detected in the conventional bilayer La$_3$Ni$_2$O$_7$-2222~\cite{La3Ni2O7-2222-anomaly} and the trilayer La$_4$Ni$_3$O$_{10}$~\cite{La4Ni3O10_anomaly_resistivity}, suggesting a possible spin density wave (SDW) transition in the 1313 La$_3$Ni$_2$O$_7$ polymorph. In La$_3$Ni$_2$O$_7$-2222, all experimental studies (neutrons, RIXS, NMR, and $\mu$SR) agree on a SDW propagation vector $q$ = (0.25, 0.25) (in pseudotetragonal notation) with two possible stripe models: a double-spin stripe or a single charge-spin stripe~\cite{2222_sdw_spin_stripe_gupta,Chen2024_rixs_327_2222,Chen_2024_PRL_muSR_La3Ni2O7_2222,neutrons, NMR}. In the trilayer La$_4$Ni$_3$O$_{10}$ neutron data were taken to test for the presence of
a SDW concomitant with the charge-density wave~\cite{interwined_charge_density_order}. The intensity
distribution of the derived superlattice reflections is consistent with a rather unusual magnetic state: outer planes that are antiferromagnetically coupled and no moment
on the inner planes. The slight
incommensurability of the SDW ordering vector results
in an approximate 5-period stripe in the plane.

The crystal structure of the La$_{3}$Ni$_{2}$O$_{7}$-1313 polymorph at ambient pressure remains ambiguous as it can reasonably be refined within different space groups due to the presence of weak reflections in the home-lab X-ray diffraction data ~\cite{1313Polymorphism202JACSmitchel,Puphal_PhysRevLett.133.146002,Wang_2024_Cmmm}.  Chen \textit{et al.} found that La$_3$Ni$_2$O$_7$-1313 crystallizes in an orthorhombic space group, with reflection conditions consistent with $Cmmm$ symmetry for most of their crystals, although they note evidence for a competing $Imma$ variant ~\cite{1313Polymorphism202JACSmitchel}.
Wang \textit{et al.} similarly concluded that this new polymorph is best described in a $Cmmm$ space group \cite{Wang_2024_Cmmm}. In contrast,  Puphal \textit{et al.} reported that an $Fmmm$ space group provides the best
refinement instead~\cite{Puphal_PhysRevLett.133.146002}. Importantly, the  $Cmmm$ structure
of ambient-pressure La$_3$Ni$_2$O$_7$-1313 (that is non-superconducting)  displays 180$^{\circ}$ Ni-O-Ni  out-of-plane bond angles in the
trilayer block, implying that this structural feature alone may be insufficient to give rise to superconductivity. Hence, it is important to understand if the $Cmmm$ structure is indeed the most feasible one for this new nickelate polymorph or if the $Imma$ or $Fmmm$ structures (that do contain octahedral tilts) are more stable instead. The electronic structure of the $Fmmm$ and $Cmmm$ phases has been scrutinized in some previous work and it seems to show some similarities and differences with the 2222 counterpart~\cite{Lecherman_PRM_1313_2024,yangzhang_PRB_2023_electronic_structure_La3Ni2O7, LaBollita_PhysRevB.110.155145,arxivheptingmono-tri}. ARPES data taken on pure La$_{3}$Ni$_{2}$O$_{7}$-1313 single crystals indicate an electronic structure for this polymorph that is analog to that of the isolated trilayer La$_4$Ni$_3$O$_{10} $\cite{ChristineC_2025_ARPES_1313}.

To shed light on the differing data reported for La$_3$Ni$_2$O$_7$-1313, here we study the structural properties and the related electronic structure of this material as a function of pressure using first-principles calculations and symmetry analysis.  We start by investigating the structural stability of the experimental structure with higher-symmetry at ambient pressure (\Cmmm) and explore the symmetry connections with the other reported space groups symmetries for this material. 
Our lattice dynamics calculations reveal that the \Cmmm phase exhibits multiple instabilities. Using group theory analysis, we find that the distortions associated with the irreducible representations (irreps) of the unstable phonon branches could transform the \Cmmm structure into different space groups, including the experimentally reported \Imma. Under pressure, we find that the tetragonal $P4/mmm$ structure becomes stable, consistent with experiments. The magnetic tendencies indicate that the electronic structure of  La$_3$Ni$_2$O$_7$-1313 at ambient pressure is dominated by the trilayer block as the single-layer is in a Mott-insulating state, consistent with ARPES \cite{ChristineC_2025_ARPES_1313}.

\section{Computational Methods}

Structural optimizations for La$_3$Ni$_2$O$_7$-1313 in the different space group symmetries and at different pressures were performed using density functional theory (DFT) non-spin-polarized calculations  with the Vienna \textit{ab-initio} Simulation Package (VASP) \cite{vasp1993, vasp1996, prbpawblochl1995}. Both the internal coordinates and the lattice parameters were relaxed using the Perdew-Burke-Ernzerhorf (PBE) version of the generalized gradient approximation (GGA)  as the exchange correlation functional \cite{pbe}. A $k$-mesh of 10$\times$10$\times$4 was used for the \Cmmm structure, while  $k$-meshes of 4$\times$4$\times$2 were used for both the \Imma and \Fmmm structures. A force convergence criterion of $10^{-4}$~eV/\AA~and an energy cutoff of 520 eV were employed in all cases.

Phonon dispersion calculations for the \Cmmm structure were carried out using Density Functional Perturbation Theory (DFPT), as implemented in the VASP-DFPT package Phonopy~\cite{phonopy}. A unit cell of size 2$\times$2$\times$1 was chosen. An energy convergence criteria of 10$^{-8}$ eV and a $k$-mesh of 4$\times$4$\times$1 was used. To analyze the unstable phonon branches and their irreps, we used the ISODISTORT module of the ISOTROPY software suite~\cite{ISOTROPY_FULL,ISODISTORT}. The character tables for each irrep were examined using the Bilbao Crystallographic Server~\cite{Bilbao_AMPLIMODES,crystallographic_point_group_Bilbao}, and the visualization of crystal structures was performed with the VESTA software~\cite{vesta_software}.

Subsequently, spin-polarized electronic-structure calculations were performed for both the \Cmmm and \Imma structures also using VASP. An energy cut-off of 400 eV and a $k$-mesh of 4$\times4\times$2 were used to sample the Brillouin zone.
To study the magnetic tendencies at ambient pressure, we employed the rotationally invariant LDA + $U$ method~\cite{dudarev_PRB_ldautype}. We analyzed the $U$ dependence of the energetics for different magnetic configurations using on-site Coulomb repulsion $U$ values of 3, 4 and 5 eV while the Hund's rule coupling value ($J$) was fixed to 0.7 eV.

\section{Experimental crystal structure} 

\begin{figure}
    \centering
    \includegraphics[width = 0.5\textwidth]{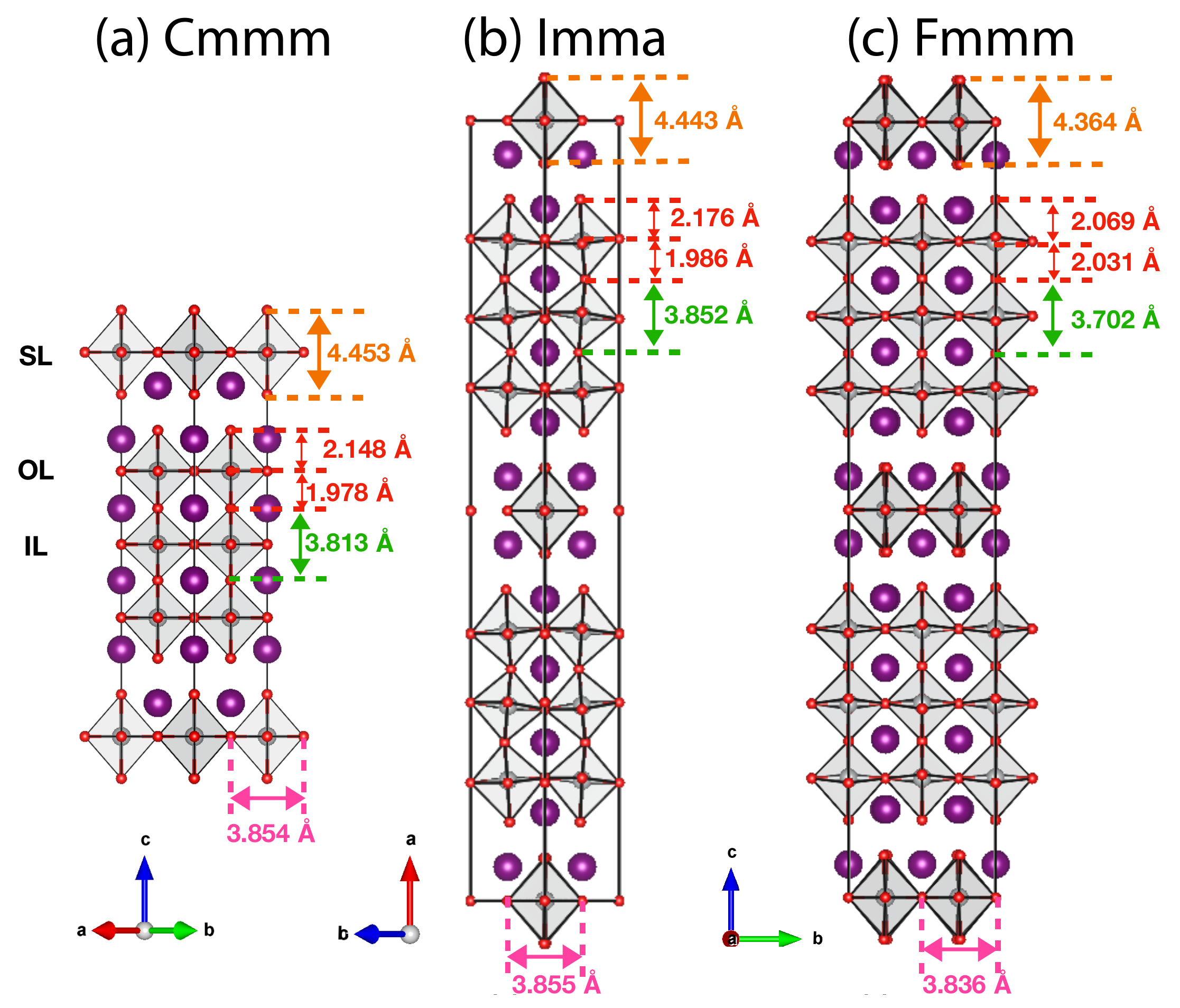}
    \caption{Crystal structure of La$_3$Ni$_2$O$_7$-1313 in the three experimentally reported space groups $Cmmm$, $Imma$, and $Fmmm$. Spheres in purple, gray, and red represent the La, Ni, and O atoms, respectively. The single-layer block is denoted by SL while the outer layer and inner layer within the trilayer are denoted as OL and IL, respectively. The in-plane and out-of-plane Ni-O bond lengths are shown explicitly.}
    \label{Fig.:crystals}
\end{figure}

The experimental crystal structure of La$_3$Ni$_2$O$_7$-1313 in the three different space groups reported ($Cmmm$, $Imma$, and $Fmmm$) is shown in Fig.~\ref{Fig.:crystals} wherein the  single-layer+trilayer structure of this polymorph can be clearly observed. 
There are three inequivalent Ni atoms in the structure: one from the single-layer block (SL) and two from the trilayer that we call inner layer (IL) and outer layer (OL). The relevant Ni-O bond lengths are highlighted in Fig.~\ref{Fig.:crystals}. The in-plane and out-of-plane Ni-O bond lengths in the $Cmmm$ and $Imma$ structures are very similar and also consistent with those reported for the individual structural components, namely, the single layer La$_2$NiO$_4$ and trilayer La$_4$Ni$_3$O$_{10}$ nickelates (see Appendix \ref{appendix:A}). In contrast, the Ni-O bond lengths of the experimentally-resolved $Fmmm$ structure can be seen to be markedly different. For both the \Imma and \Fmmm structures, the trilayer block exhibits clear octahedral tilts with a Ni-O-Ni out-of-plane angle of 166.48$^{\circ}$ for \Imma and 171.42$^{\circ}$ for \Fmmm.
Due to the out-of-plane NiO$_{6}$ octahedral tilts, the \Imma and \Fmmm unit cells are doubled along the out-of-plane direction compared to the \Cmmm structure where, as mentioned above, the Ni-O-Ni bond angle across the apical oxygens is 180$^\circ$. The NiO$_{6}$ octahedra in the single-layer block remain untilted across all three space groups.

\section{Results}

\subsection{Structural Stability}

\begin{table}
    \centering
     \setlength{\tabcolsep}{3pt} 
      \caption{Irreducible representations (irreps), frequencies, order parameter directions (OPDs) and the corresponding transformed structures for the lowest-frequency unstable phonon modes at the Y and T points for \Cmmm-La$_3$Ni$_2$O$_7$-1313.}
      \vspace{4pt}
    \begin{tabular}{ c  c  c  c}
     \hline \hline
     \multicolumn{1}{c}{k-points}&
      \multicolumn{1}{c}{$\text{Frequency (meV)}$} & 
      \multicolumn{1}{c}{OPD} & 
       \multicolumn{1}{c}{Structure}\\
      \hline 
      
      Y & Y$_{2}^{+}$ ; 18.28$i$  & (a) & Pbam \\  
      & Y$_{3}^{+}$ ; 14.00$i$   & (a) &Pmna  \\
      & Y$_{4}^{+}$ ; 13.66$i$  & (a) &Pmna  \\
      
      \hline
      T & T$_{1}^{-}$ ; 18.44$i$  & (a) &Ibam  \\
      & T$_{4}^{-}$ ; 13.89$i$  & (a) &Imma  \\
      & T$_{3}^{-}$ ; 13.56$i$ & (a) &Imma \\
      \hline
      \end{tabular}
      \label{table:irreps_Cmmm}
\end{table}

\begin{figure*}
    \centering
    \includegraphics[width = 1.0\textwidth]{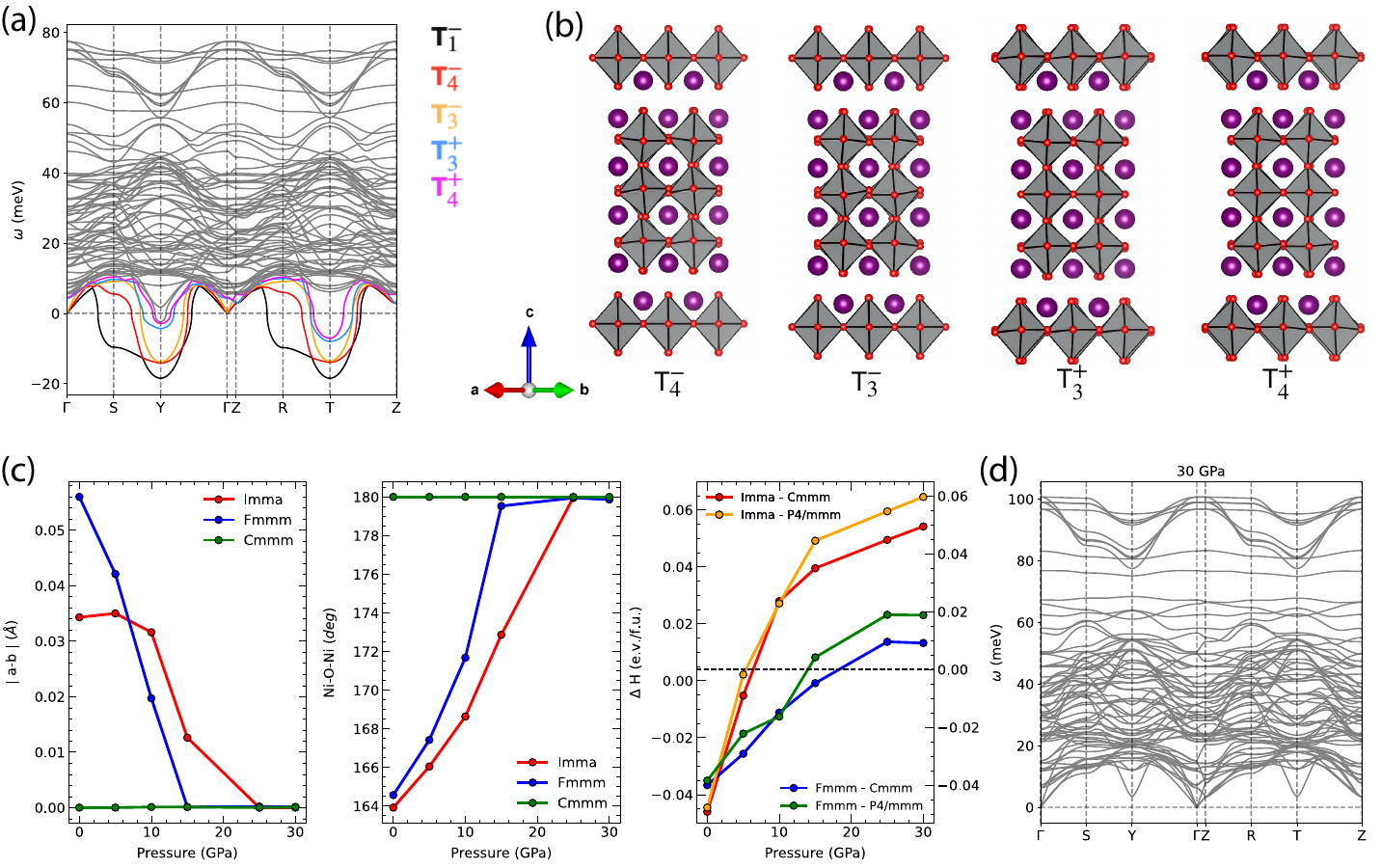}
    \caption{(a) Phonon dispersion of \Cmmm-La$_3$Ni$_2$O$_7$-1313 at ambient pressure. The different unstable modes at various high symmetry points of the \Cmmm Brillouin zone are highlighted with different colors. The high symmetry points correspond to: $\Gamma =$ (0.0, 0.0, 0.0), S $=$ (0.0, 0.5, 0.0), Y $=$ (0.5, 0.5, 0.0), Z $=$ (0.0, 0.0, 0.5), R $=$ (0.0, 0.5, 0.5), T $=$ (0.5, 0.5, 0.5). (b)  Distortions associated with each of the four irreps of the unstable phonon branches at T that lead to an $Imma$ structure: T$_{4}^{-}$,  T$_{3}^{-}$, T$_{3}^{+}$, and T$_{4}^{+}$. (c) (left) Evolution of the lattice parameters of the \Cmmm, \Imma, and \Fmmm structures with pressure. (middle) Evolution of the Ni-O-Ni bond angle across the apical oxygen with pressure for the three different space groups. (right) Enthalpy (H = E + PV) evolution of \Imma and \Fmmm structures with respect to the \Cmmm (left y-scale) and \textit{P4/mmm} (right y-scale) structures. At ambient pressure the \Imma structure has the lowest enthalpy. The $P4/mmm$ space group becomes more stable at $\sim$ 13 GPa, consistent with experiments. (d) Phonon dispersion of the $P4/mmm$ structure at 30 GPa that is dynamically stable. All the first-principles results in this figure correspond to fully relaxed structures and non-spin-polarized calculations.}
    \label{Fig.:phonon_Cmmm_Imma_Fmmm}
\end{figure*}

We start by analyzing the dynamical stability of the crystal structure with higher symmetry at ambient pressure for La$_3$Ni$_2$O$_7$-1313 ($Cmmm$) by studying its phonon dispersion, shown in Fig.~\ref{Fig.:phonon_Cmmm_Imma_Fmmm}(a). We perform our phonon analysis for La$_3$Ni$_2$O$_7$-1313 in the nonmagnetic state as analog calculations in La$_3$Ni$_2$O$_7$-2222 and  La$_4$Ni$_3$O$_{10}$ give rise to stable phonon dispersions in their respective experimentally reported  structures with $Amam$ and $P2_{1}/c$ symmetries ~\cite{labollita2024trilayer, phonons_4310}. In contrast, the calculated phonon dispersion for the $Cmmm$ structure of La$_3$Ni$_2$O$_7$-1313 exhibits several instabilities, the lower ones in energy being along the Brillouin zone edge Y-T (see Fig.~\ref{Fig.:phonon_Cmmm_Imma_Fmmm}(a)). 

The largest instability in the phonon dispersion corresponds to a nondegenerate branch with irreps Y$_{2}^{+}$ and T$_{1}^{-}$ at the Y and T points, respectively. Subsequently, an almost doubly-degenerate branch can be observed with irreps Y$_{3}^{+}$, Y$_{4}^{+}$  and  T$_{4}^{-}$, T$_{3}^{-}$. These branches are almost dispersionless along Y-T. Higher in energy lies another quasi-degenerate branch (with stronger dispersion along Y-T) with irreps Y$_{3}^{-}$, Y$_{4}^{-}$ and  T$_{3}^{+}$, T$_{4}^{+}$. The distortion patterns of each irrep at the T point of the Brillouin zone are shown in  Fig.~\ref{Fig.:phonon_Cmmm_Imma_Fmmm}(b) (the analog distortions at the Y point are shown in Appendix \ref{appendix:C}). The distortion due to the T$_{1}^{-}$ irrep corresponds to an in-plane rotation of oxygen atoms of the inner layer of the trilayer block. 
 The T$_{4}^{-}$  and T$_{3}^{-}$ modes mainly involve tilting of the NiO$_{6}$ octahedra of the trilayer blocks about the [100] and [010] axis respectively, without introducing any displacements in the single-layer block. As such, the T$_{4}^{-}$ and T$_{3}^{-}$ distortions lead to a Ni-O-Ni apical bond angle different from 180$^{\circ}$. The mode with T$_{3}^{+}$ irrep involves an out-of-plane rotation of the NiO$_{6}$ octahedra about the [100] axis while the T$_{4}^{+}$ involves a rotation about the [010] axis. In addition to the NiO$_{6}$ octahedral rotation of the trilayer block, the modes with T$_{3}^{+}$ and T$_{4}^{+}$ irreps also lead to the rotation of the NiO$_{6}$ octahedra of the single-layer block about the [100] axis for T$_{3}^{+}$ and about the [010] axis for T$_{4}^{+}$.

We now focus on the imaginary frequencies of the lowest-frequency phonon modes at Y and T that are listed in Table~\ref{table:irreps_Cmmm} (details on the rest of the modes are shown in Appendix \ref{appendix:C}). 
Table~\ref{table:irreps_Cmmm} also shows the space group symmetry of the structures obtained using the distortions associated with each of the irreps of these most unstable phonon modes for the  $Cmmm$ structure at both Y and T.  It is important to note that the distortions associated to the irreps at the Y point never lead to crystal structures with \Fmmm or \Imma symmetry (the other two experimentally reported space groups). In contrast, the distortions associated with the T$_{4}^{-}$ and T$_{3}^{-}$ irreps can lead to a structure with $Imma$ symmetry. The distortions associated with T$_{3}^{+}$ and T$_{4}^{+}$ irreps also lead to an $Imma$ structure, even though these do not correspond to the lowest-frequency modes.

After studying its dynamical stability at ambient pressure, we now make enthalpy considerations for La$_3$Ni$_2$O$_7$-1313 in relation to the proposed space group symmetries. We note that enthalpy calculations from first-principles in the nonmagnetic state for La$_3$Ni$_2$O$_7$-2222 and La$_4$Ni$_3$O$_{10}$ have revealed the transition from an orthorhombic ($Amam$) or monoclinic ($P2_{1}/c$)
crystal setting to a tetragonal ($I4/mmm$) structure under pressure~\cite{labollita2024trilayer, zhang2024structural}, in agreement with XRD data~\cite{wang2024structure,Li_2025_acs_xrd_4310,wang_2025_inorganic_chemsitry_2222_structural_phase_transition}. Analog calculations in La$_3$Ni$_2$O$_7$-1313 (see Fig.~\ref{Fig.:phonon_Cmmm_Imma_Fmmm}(c)) reveal the same type of transition with a \textit{P4/mmm} structure  quickly becoming more stable once pressure is applied (using any experimentally reported space group symmetry as a starting point), also in agreement with experiments \cite{Puphal_PhysRevLett.133.146002}. Across the transition, the difference between the $a$ and $b$ lattice constants becomes zero, leading to a tetragonalized crystal structure at $\sim$ 15 GPa for an \Fmmm and at 25 GPa for an \Imma space group as a starting point. Concomitantly, the Ni-O-Ni apical bond angles straighten to 180$^{\circ}$. 
This finding is consistent with the phonon calculations that show that, with increasing pressure, the soft phonon branches harden leading to a dynamically stable crystal structure with \textit{P4/mmm} symmetry at $\sim$ 30 GPa (see Fig.~\ref{Fig.:phonon_Cmmm_Imma_Fmmm}(d) and a full pressure evolution in Appendix~\ref{appendix:B}). Importantly, the enthalpy results at ambient pressure seem to reinforce the likelihood of a structure containing octahedral tilts at ambient pressure in La$_3$Ni$_2$O$_7$-1313. At 0 GPa, the \Imma is the most stable structure, with an energy approximately 40 meV/f.u. lower than that of the \Cmmm structure. The \Fmmm structure is also more stable than the \Cmmm one but it is less stable than the \Imma.

\begin{figure*}
   \centering
     \includegraphics[width=1.0\linewidth]{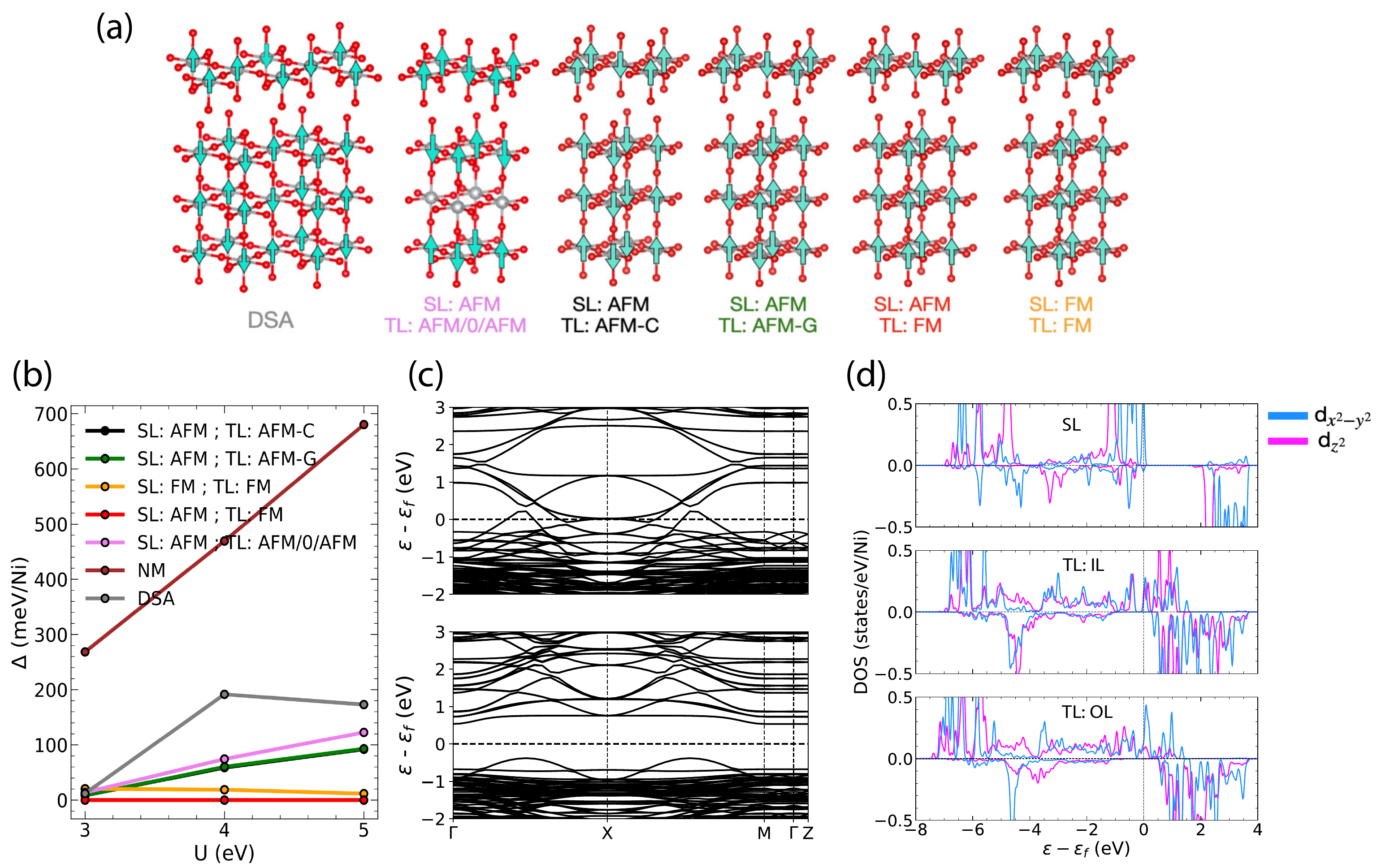}
   \caption{Magnetic tendencies of \Imma-La$_3$Ni$_2$O$_7$-1313 at ambient pressure. (a) Spin configurations analyzed in this work. SL denotes the single-layer, TL the trilayer Ni atoms. (b) Energetics of the different magnetic configurations as a function of $U$.  (c) Band structure for the ground state configuration (SL:AFM; TL:FM) at $U$ = 4 eV in both the majority and minority spin channel. (d) Orbitally ($e_g$) projected density of states for SL, TL:IL (inner-layer) and TL:OL (outer-layer) Ni atoms in the magnetic ground state ($d_{z^2}$ pink, $d_{x^2-y^2}$ blue). }
    \label{fig:Imma_band_dos_moments}
\end{figure*}

To summarize, our analysis of the unstable phonon branches in the experimentally reported structure with higher-symmetry at ambient pressure for La$_3$Ni$_2$O$_7$-1313 ($Cmmm$) shows that this structure can transform into one of the other experimentally reported space groups (\Imma) through the distortions associated to irreps at the T point (an $Fmmm$ structure cannot be obtained from any of the lowest-frequency modes). Enthalpy considerations also seem to favor an \Imma space group symmetry. Therefore, based on our analysis, the \Imma structure seems to be the most feasible one out of the three experimentally reported structures for La$_3$Ni$_2$O$_7$-1313, indicating that this hybrid phase (akin to the conventional bilayer and trilayer compounds) likely contains octahedral tilts.

Some final remarks about the structural stability of La$_3$Ni$_2$O$_7$-1313 are in order. 
 While based on our symmetry analysis and on energetics, the \Imma structure seems to be the most feasible one out of the three experimentally reported ones, it is important to note that the $Imma$ structure involves condensation of only the quasi-doubly-degenerate branch at T in the phonon dispersion shown in Fig.~\ref{Fig.:phonon_Cmmm_Imma_Fmmm}(a). Using not only individual irreps of the unstable phonon modes but also their sum with each other, other structures can be generated that involve condensation of not only the quasi-doubly-degenerate branch but also the nondegenerate (most unstable) branch, as done in Ref.~\cite{subedi2024pressuretunablestructuralinstabilitiessinglelayertrilayer}. However, we have found that such structures (for example \textit{Pnma}) do not provide an energy gain. Further, while it would be useful to perform similar structural stability studies in spin-polarized calculations (as done for La$_3$Ni$_2$O$_7$-2222~\cite{metal_insulator_transition_La3Ni2O7_2024}), the lack of information about the $q$ of the density wave state in the single-layer trilayer polymorph prevents us from doing such an analysis in a meaningful manner at this point.

\subsection{Electronic structure and magnetic properties}


\vspace{0.1in}

After achieving a basic understanding of the structural stability of La$_3$Ni$_2$O$_7$-1313 (that points to the presence of octahedral tilts), we now explore magnetic tendencies in this material to scrutinize its low-energy electronic structure. In the nonmagnetic state at the DFT level (as reported by us and others~\cite{yangzhang_PRB_2023_electronic_structure_La3Ni2O7,1313Polymorphism202JACSmitchel}), the single-layer bands cross the Fermi level. This result disagrees with recent ARPES experiments~\cite{ChristineC_2025_ARPES_1313} that find a Fermi surface analog to that of the trilayer nickelate La$_4$Ni$_3$O$_{10}$, with the single-layer bands being out of the picture. To shed light on this issue, we performed LDA+$U$ calculations for different magnetic configurations for a range of $U$s from 3 to 5 eV in both $Cmmm$ and $Imma$ space group symmetries.   The magnetic configurations we attempted (depicted in Fig. \ref{fig:Imma_band_dos_moments}(a)) are: a) a double spin-stripe (DSA), analog of the magnetic ground state of La$_3$Ni$_2$O$_7$-2222. b) A checkerboard antiferromagnetic configuration in the single-layer and an antiferromagnetic-C-type configuration in the trilayer (SL: AFM; TL: AFM-C). c) A checkerboard antiferromagnetic state in the single-layer and an antiferromagnetic-G-type configuration in the trilayer (SL: AFM; TL: AFM-G). d) A ferromagnetic single-layer and ferromagnetic trilayer (SL: FM; TL: FM). e) A checkerboard antiferromagnetic configuration in the single-layer and a ferromagnetic configuration in the trilayer (SL: AFM; TL: FM). f) A checkerboard AFM configuration in the single-layer as well as in the outer-layers of the trilayer, with the inner-layer of the trilayer having zero moments (SL: AFM TL: AFM/0/AFM). This latter magnetic state, with a node in the inner layer of the trilayer and antiferromagnetically coupled outer layers (in agreement with neutron data ~\cite{interwined_charge_density_order}) has been shown to be the magnetic ground state for the trilayer nickelate La$_4$Ni$_3$O$_{10}$ from DFT for a $\sqrt{2}\times\sqrt{2}$-sized cell~\cite{labollita2024trilayer, dagotto_4310}.  While we analyze all of these magnetic states in $Cmmm$ and $Imma$ space group symmetries, we present in the main text the results for the $Imma$ structure only (the $Cmmm$ results are shown in Appendix~\ref{appendix:D}). Regardless of the space group chosen, the SL: AFM; TL: FM is the magnetic ground state for all $U$ values analyzed here for the magnetic states and cell sizes we analyze (see Fig.~\ref{fig:Imma_band_dos_moments}(b) and Appendix~\ref{appendix:D}). We emphasize once again that La$_3$Ni$_2$O$_7$-1313 has a density-wave instability as reflected in the ambient-pressure transport data~\cite{1313Polymorphism202JACSmitchel, 1313_anomaly} but information on the nature of the density-wave state and its $q$ would be needed in order to further scrutinize that possibility from first-principles in appropriate/larger supercells.

The electronic structure for the magnetic ground state we obtain for La$_3$Ni$_2$O$_7$-1313  is consistent with ARPES data in terms of the dominance of the trilayer block. Fig.~\ref{fig:Imma_band_dos_moments} (c, d) shows the electronic band structure and $e_g$-density of states for this SL: AFM; TL: FM configuration at $U$ = 4 eV. As expected, given the average $d^{7.5}$ filling for the Ni atoms, orbitals of $d_{z^2}$ and $d_{x^2-y^2}$ character dominate the electronic structure around the Fermi level, with most of the weight at the Fermi energy coming from the planar $d_{x^2-y^2}$ orbitals.  Importantly, the DOS of the single-layer Ni is clearly gapped as a consequence of the AFM order in the plane (and the inclusion of an on-site Coulomb repulsion), in line with the Mott insulating state of La$_2$NiO$_4$~\cite{GWGuo_1988_La2NiO4}. As a consequence, only states from the trilayer Ni atoms are relevant at the Fermi level. The derived magnetic moments in the magnetic ground state at $U$= 4 eV are $\sim$ 1.5 $\mu_B$ for the single layer nickel, $\sim$ 1.4 $\mu_B$ for the outer layer Ni of the trilayer and a reduced $\sim$ 1.2 $\mu_B$ for the inner nickel of the trilayer block (further information on the evolution of the magnetic moments can be seen in Appendix~\ref{appendix:D}). These derived moments are consistent with a higher $d$ occupation for the single-layer Ni atoms. The moments disproportionate within the trilayer block, indicating a different effective doping for inner and outer planes, as shown in previous work for La$_4$Ni$_3$O$_{10}$~\cite{labollita2024trilayer}. The electronic structure we just described for La$_3$Ni$_2$O$_7$-1313 is in agreement with DFT+DMFT calculations in La$_3$Ni$_2$O$_7$-1313 that show that when dynamical correlations are taken into account, the single-layer block is in a Mott-insulating regime~\cite{LaBollita_PhysRevB.110.155145}. More importantly, our results, as mentioned above, are in accordance with recent ARPES data that show that the low-energy physics of this material is very similar to that of the `isolated' trilayer La$_4$Ni$_3$O$_{10}$ system~\cite{ChristineC_2025_ARPES_1313}. Indeed, the unfolded band structure of our magnetic ground state is very similar to that of the trilayer material~\cite{li2017}, as shown in Appendix~\ref{appendix:E}. As the electronic structure at high-pressure in $P4/mmm$ symmetry has been intensively studied in previous work \cite{Lecherman_PRM_1313_2024,yangzhang_PRB_2023_electronic_structure_La3Ni2O7, LaBollita_PhysRevB.110.155145,arxivheptingmono-tri}, we do not analyze it here.

\section{Conclusions}

In summary, we have employed first-principles methods and group theory analysis to study the structure and electronic structure of the newly discovered single-layer-trilayer La$_{3}$Ni$_{2}$O$_{7}$ polymorph. At ambient pressure, the structure of this material in the experimentally reported space group with higher symmetry ($Cmmm$) exhibits multiple unstable phonon branches. Distortions associated with these instabilities lead to one of the other
experimentally reported space groups ($Imma$), that is the most stable space group at ambient pressure also from enthalpy considerations. Importantly, the $Imma$ structure contains octahedral tilts, indicating that the hybrid La$_{3}$Ni$_{2}$O$_{7}$-1313 polymorph is likely structurally similar to the conventional bilayer and trilayer RP phases. Under pressure, a $P4/mmm$ structure becomes stable, in agreement with XRD experiments. The magnetic tendencies at ambient pressure show that the electronic structure of La$_{3}$Ni$_{2}$O$_{7}$-1313 in its magnetic ground state is dominated by the trilayer block as the single-layer is
in a Mott insulating regime. This is in agreement with recent ARPES data.  Our results beg for a more detailed analysis of the ambient pressure structure of La$_{3}$Ni$_{2}$O$_{7}$-1313 from synchrotron or neutron diffraction to clarify the presence of octahedral tilts. Further experiments to determine the nature (and propagation vector) of the apparent density-wave state should also be pursued to clarify the magnetic ground state of this material and its connection to the conventional bilayer and trilayer RPs.

\section*{acknowledgements} 

We acknowledge NSF grant No. DMR-2045826 and the ASU research computing center for HPC resources.



%

\appendix

\onecolumngrid

\newpage
\appendix

\section{\label{appendix:A} Structural data for La$_2$NiO$_4$ and La$_4$Ni$_3$O$_{10}$}

\begin{figure*}[h!]
    \centering
    \includegraphics[width=0.6\linewidth]{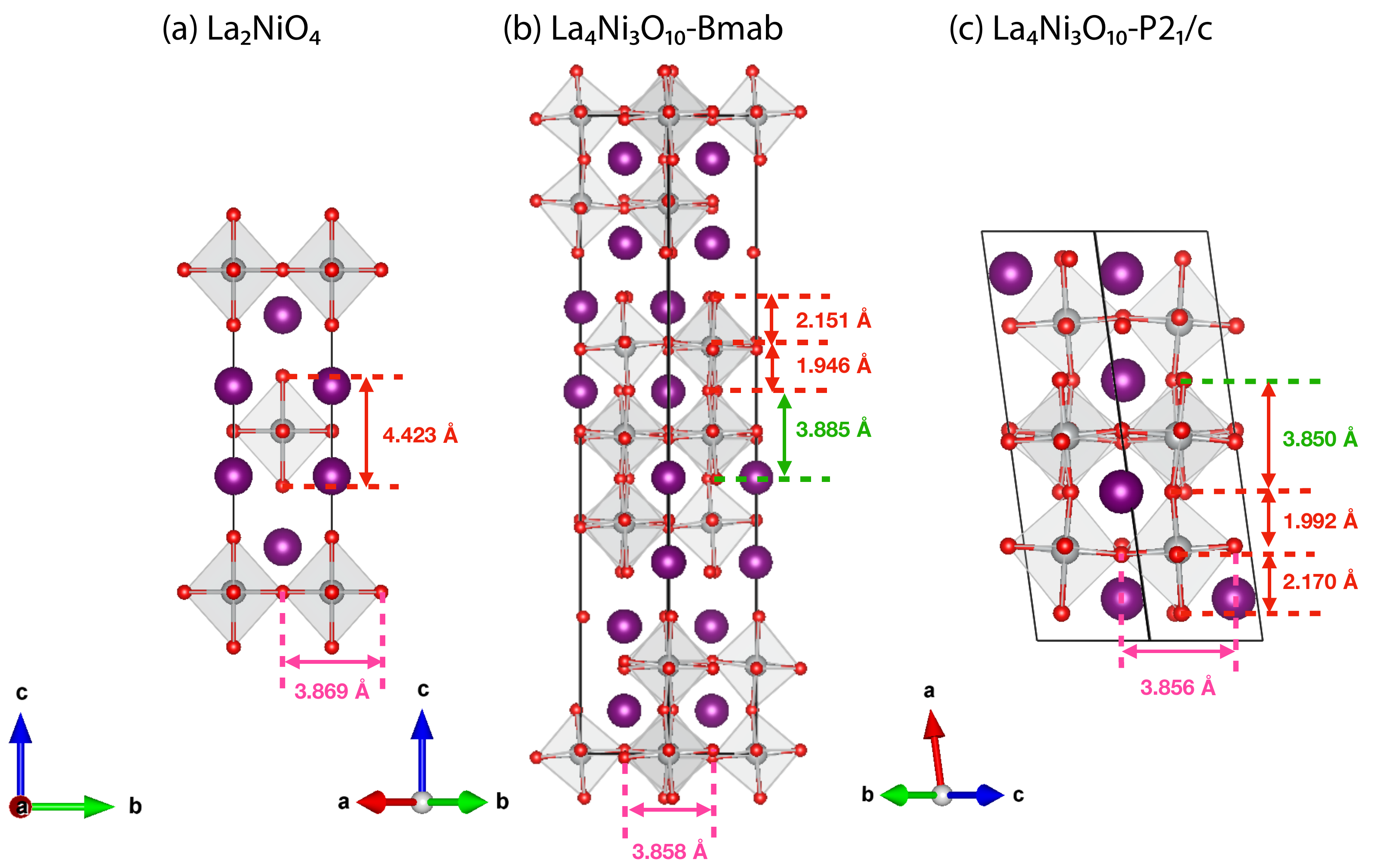}
    \caption{Structures of (a) La$_2$NiO$_4$ and of La$_4$Ni$_3$O$_{10}$ in both $Bmab$ (b) and $P2_1/c$ (c) symmetries. Relevant bond lengths as obtained from experimental structural data are shown.}
    \label{fig:mono_tri}
\end{figure*}

Fig.~\ref{fig:mono_tri} shows the structural data for La$_2$NiO$_4$ and La$_4$Ni$_3$O$_{10}$ extracted from Refs.~\cite{SKINNER_214_lattice,Junjie_Zhang_4310_2020} in order to compare the relevant bond lengths of the two individual structural components with those of La$_{3}$Ni$_{2}$O$_{7}$-1313 shown in Fig. \ref{Fig.:crystals}. 

\section{\label{appendix:C} Further details on the unstable phonon branches for the \Cmmm structure} 

\begin{figure*}
    \centering
    \includegraphics[width=\linewidth]{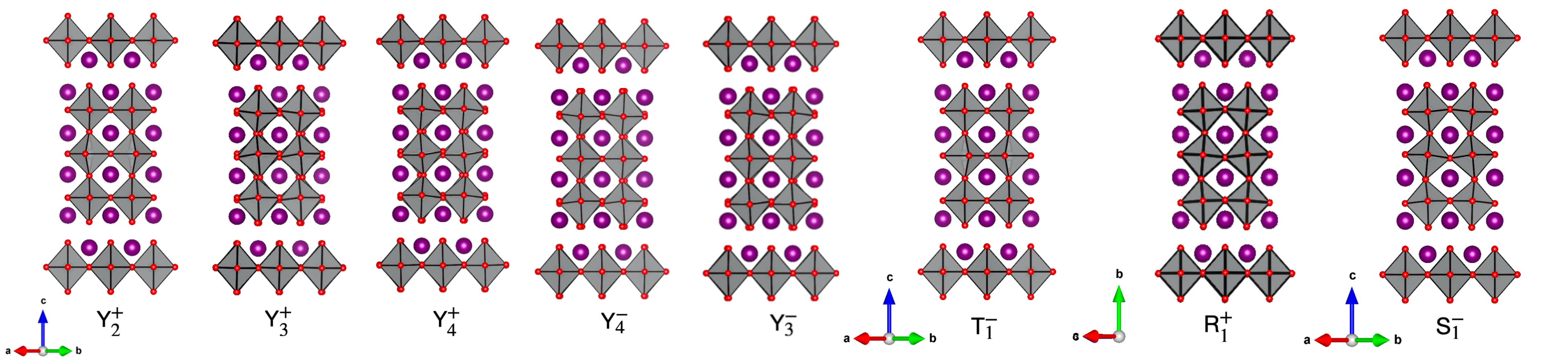}
    \caption{Distortions corresponding to the irreps of the unstable phonon branches of $Cmmm$-La$_{3}$Ni$_{2}$O$_{7}$-1313 at the Y, T, R, and S points of the Brillouin zone not shown in the main text.}
    \label{fig:distortions_S_Y_R}
\end{figure*}

Table \ref{table:irreps_complete} shows a complete list of irreps, frequencies, OPDs and corresponding distorted structures for all unstable phonon modes of \textit{Cmmm}-La$_{3}$Ni$_{2}$O$_{7}$-1313. 
Fig.~\ref{fig:distortions_S_Y_R} shows the displacement patterns of high-symmetry modes at the Y and T points of the Brillouin zone, as described in the main text, as well as at the S and R points. At the S point, the distortion mode corresponds to S$_{1}^{-}$ whose displacement patterns lead to the tilting of NiO$_{6}$ octahedra of the trilayer block about both the [100] and [010] in-plane axis. The octahedra of the single-layer block remain undistorted. At the R point, the unstable normal mode has irrep R$_{1}^{+}$. Similarly to S$_{1}^{-}$, the displacement patterns lead to the rotation of the trilayer block about both the [100] and [010] axis which ultimately leads to the out-of-plane tilting of NiO$_{6}$ octahedra. Note that via distortions at R, an $Fmmm$ structure could be obtained. 

\begin{table}
    \centering
     \setlength{\tabcolsep}{3pt} 
      \caption{Complete set of irreducible representations (irreps), frequencies, order parameter directions (OPDs) and the corresponding transformed structures for all unstable phonon modes for \Cmmm-La$_3$Ni$_2$O$_7$-1313.}
      \vspace{4pt}
    \begin{tabular}{ c  c  c  c c c c c c}
     \hline \hline
     \multicolumn{1}{c}{k-point}&
      \multicolumn{1}{c}{Frequency (meV)} & 
      \multicolumn{1}{c}{OPD} & 
       \multicolumn{1}{c}{Structure}&
          \multicolumn{1}{c}{k-point}&
      \multicolumn{1}{c}{$\text{Frequency (meV)}$} & 
      \multicolumn{1}{c}{OPD} & 
       \multicolumn{1}{c}{Structure}
       \\
      \hline 
       S & S$_{1}^{-}$ ; 9.58$i$ & (a,0), (a,b)  & P2/c &  R & R$_{1}^{+}$ ; 9.65$i$  & (a,0), (a,b)  &C2/m \\ 
       & & & & & & (a,a) & Fmmm   \\ 

      
      Y & Y$_{2}^{+}$ ; 18.28$i$  & (a) & Pbam &  T & T$_{1}^{-}$ ; 18.44$i$  & (a) &Ibam\\  
      & Y$_{3}^{+}$ ; 14.00$i$, 2.83$i$  & (a) &Pmna &  & T$_{4}^{-}$ ; 13.89$i$  & (a) &Imma \\
      & Y$_{4}^{+}$ ; 13.66$i$  & (a) &Pmna &  & T$_{3}^{-}$ ; 13.56$i$ & (a) &Imma  \\
      &Y$_{3}^{-}$ ; 2.35$i$  & (a) &Pmma &  & T$_{3}^{+}$ ; 7.94$i$  & (a) &Imma \\
      &Y$_{4}^{-}$ ; 4.27$i$  & (a) &Pmma &  & T$_{4}^{+}$ ; 7.07$i$  & (a) &Imma \\

      \hline
      \end{tabular}
      \label{table:irreps_complete}
\end{table}

\section{\label{appendix:B} Phonon dispersions in P4/mmm symmetry} 

\begin{figure*}[h!]
    \centering 
    \includegraphics[width = 0.85\linewidth]{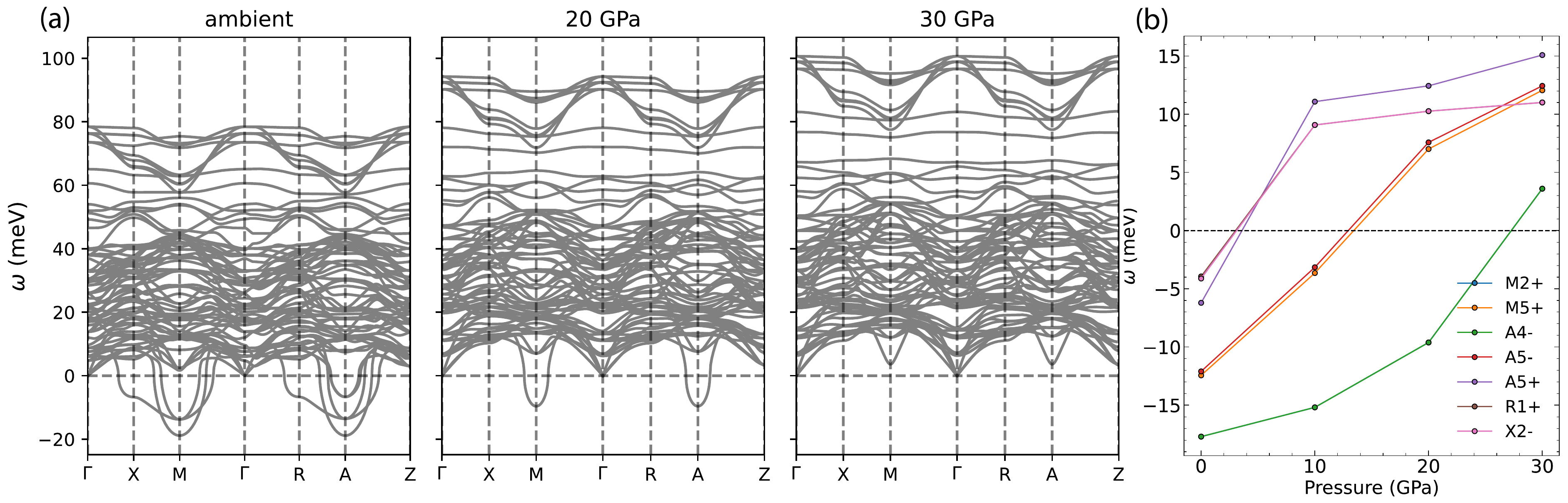}
    \caption{(a) Evolution of the phonon dispersion of \textit{P4/mmm}-La$_{3}$Ni$_{2}$O$_{7}$-1313 with pressure. (Left panel) ambient pressure, (middle panel) 20 GPa, (right panel) 30 GPa. (b) Evolution of different distortion modes with pressure. M$_{2}^{+}$, A$_{4}^{-}$ and R$_{1}^{+}$, X$_{2}^{-}$ are degenerate and their plots coincide.}
    \label{fig:p4mmm_phonon_dispersion}
\end{figure*}

Fig. \ref{fig:p4mmm_phonon_dispersion} shows the evolution with pressure of the phonon dispersion of La$_{3}$Ni$_{2}$O$_{7}$-1313 in $P4/mmm$ symmetry. $Fmmm$,
$Cmmm$, and $Imma$ structures can be obtained from the $P4/mmm$ phase via
atomic displacements associated with the irreps of different unstable phonon modes (see Table \ref{table:irreps_P4mmm}). Similar to the $Cmmm$ structure, at ambient pressure, there are two unstable phonon branches at M (the most unstable branch being nondegenerate and the other being doubly degenerate). At A there is an extra doubly degenerate phonon branch. As shown in Table \ref{table:irreps_P4mmm}, the most (nondegenerate) unstable
branch has irreps
M$_{2}^{+}$ and A$_{4}^{-}$ at M and A, respectively. The subsequent doubly degenerate branch has irreps M$_{5}^{+}$ and A$_{5}^{-}$. The following doubly-degenerate branch at A has the same irrep.  The M$_{2}^{+}$ (and A$_{4}^{-}$) modes 
involve predominantly an in-plane rotation of the middle layer of the NiO$_6$
octahedra within the trilayer. The M$_{5}^{+}$ (and A$_{5}^{-}$) modes involve rotation of all the NiO$_6$ octahedra in planes parallel to the
$c$ axis, similar to the leading instabilities of the $Cmmm$ phase described in the main text.   With increasing pressure, the unstable branches are gradually suppressed, as shown in Fig. \ref{fig:p4mmm_phonon_dispersion}.

\begin{table}[h!]
    \centering
     \setlength{\tabcolsep}{10pt} 
     \caption{Irreducible representations (irreps), frequencies, order parameter directions (OPDs) and the corresponding transformed structures of the two lowest-frequency unstable phonon branches at the M and A points of the Brillouin zone for \textit{P4/mmm}-La$_3$Ni$_2$O$_7$-1313.}
     \vspace{4pt}
    \begin{tabular}{ c  c  c  c}
     \hline \hline
     \multicolumn{1}{c}{k-points}&
      \multicolumn{1}{c}{$\text{Frequency (meV)}$} & 
      \multicolumn{1}{c}{$\text{OPD}$}&
       \multicolumn{1}{c}{Structure}\\
      \hline 
      
      M & M$_{2}^{+}$ ; 17.66$i$  & (a) & P4/mbm \\  
      &M$_{5}^{+}$ ; 12.70$i$ &(a,0) &Pmna \\
      & & (a,a) & Cmma \\
      & & (a,b) & P$_{2}/c$ \\
      
      \hline
      A & A$_{4}^{-}$ ; 17.70$i$  & (a)  & I4/mcm \\
      & A$_{5}^{-}$ ; 12.09$i$  & (a,0) & Imma  \\
      & & (a,a) & Fmmm \\
      & & (a,b) & C2/m \\
      \hline
      \end{tabular}
      \label{table:irreps_P4mmm}
\end{table}

\section{\label{appendix:D} Further details of the spin-polarized calculations}

\begin{figure}[h!]
    \centering
    \includegraphics[width=0.3\linewidth]{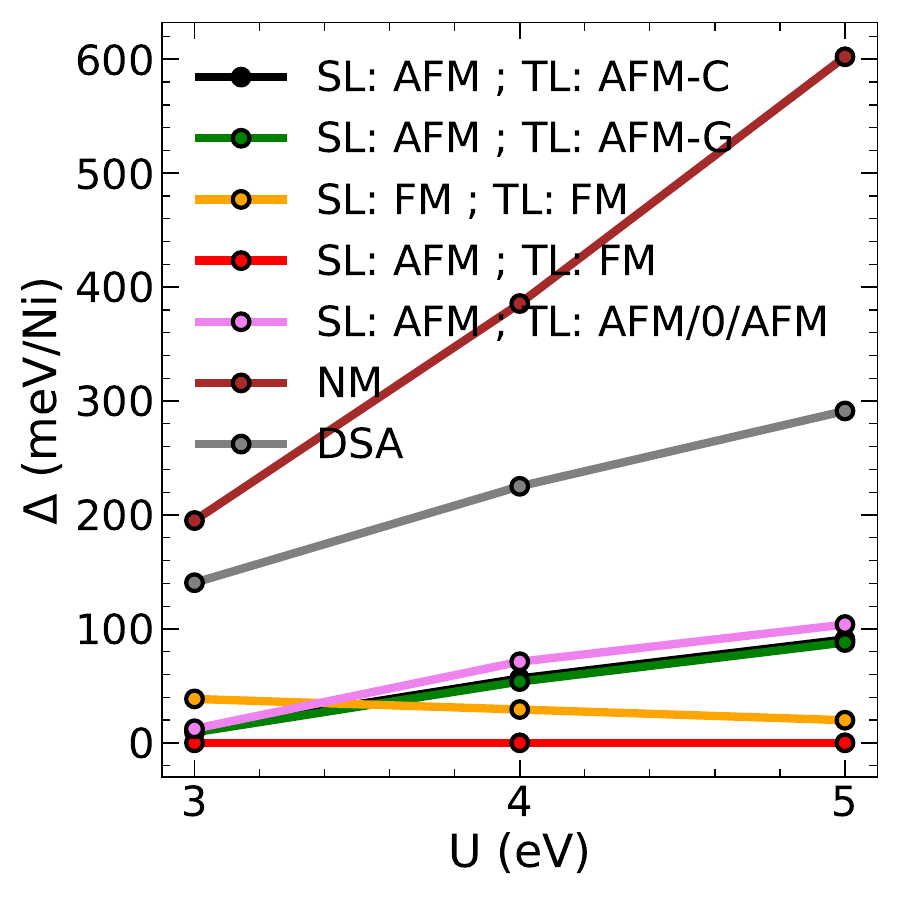}
    \caption{Energetics of different magnetic configurations depicted in the main text for \Cmmm-La$_{3}$Ni$_{2}$O$_{7}$-1313 as a function of $U$.}
    \label{fig:Cmmm_magnetic_energies}
\end{figure}

The magnetic moments obtained for each of the magnetic configurations described in the main text (for $U$ = 3, 4 and 5 eV) both for $Cmmm$ and $Imma$ symmetry are shown in Tables ~\ref{table:moments_Imma} and ~\ref{table:moments_Cmmm}. The energetics for different magnetic configurations for a $Cmmm$ structure (analog to those shown in the main text for $Imma$ symmetry) can be seen in Fig. \ref{fig:Cmmm_magnetic_energies}. The magnetic ground state in $Cmmm$ symmetry is identical (SL: AFM, TL:FM).

\begin{table}
    \setlength{\tabcolsep}{6pt}
    \renewcommand{\arraystretch}{1.2}
    \caption{ Magnetic moments of the inequivalent Ni atoms (in $\mu_B$) for $Imma$-La$_{3}$Ni$_{2}$O$_{7}$-1313 for the different magnetic states depicted in Fig. 3 as a function of $U$. SL denotes the single-layer Ni atoms, OL the outer-layer Ni atoms within the trilayer, and IL the inner-layer Ni atoms within the trilayer.}
    \vspace{4pt}
    \begin{tabular}{c c c c c c c}
        \hline\hline
           $U$ (eV) & \textcolor{gray}{DSA} &
        \textcolor{magenta}{SL: AFM } &
        SL: AFM  &
        \textcolor{green}{SL: AFM} &
        \textcolor{red}{SL: AFM } &
        \textcolor{yellow}{ SL: FM} \\
                &  &
        \textcolor{magenta}{TL: AFM/0/AFM} &
         TL: AFM-C &
        \textcolor{green}{TL: AFM-G} &
        \textcolor{red}{TL: FM} &
        \textcolor{yellow}{ TL: FM} \\
        \hline
        \multirow{3}{*}{3}
        & SL: 0.430/ -0.430  & SL: 1.395/ -1.395 & SL: 1.370/ -1.370 & SL: 1.359/ -1.359 & SL: 1.400/ -1.424 & SL: 1.532 \\
        & OL: 1.195/ -1.195  & OL: 1.093/ -1.093 & OL: 1.109/ -1.109 & OL: 1.103/ -1.103 & OL: 1.323        & OL: 1.367 \\
        & IL: 0.586/ -0.586  & IL: 0.000         & IL: 0.507/ -0.507 & IL: 0.505/ -0.505 & IL: 1.020        & IL: 1.024 \\
        \hline
        \multirow{3}{*}{4}
        & SL: 0.314/ -0.314  & SL: 1.478/ -1.478 & SL: 1.457/ -1.457 & SL: 1.448/ -1.448 & SL: 1.469/ -1.455 & SL: 1.579 \\
        & OL: 1.297/ -1.297  & OL: 1.240/ -1.240 & OL: 1.240/ -1.240 & OL: 1.242/ -1.242 & OL: 1.385        & OL: 1.426 \\
        & IL: 0.746/ -0.746  & IL: 0.000         & IL: 0.669/ -0.669 & IL: 0.676/ -0.676 & IL: 1.098        & IL: 1.096 \\
        \hline
        \multirow{3}{*}{5}
        & SL: 0.198/ -0.198  & SL: 1.549/ -1.549 & SL: 1.533/ -1.533 & SL: 1.528/ -1.528 & SL: 1.569/ -1.571 & SL: 1.625 \\
        & OL: 1.391/ -1.391  & OL: 1.367/ -1.367 & OL: 1.352/ -1.352 & OL: 1.357/ -1.357 & OL: 1.445        & OL: 1.485 \\
        & IL: 0.892/ -0.892  & IL: 0.000         & IL: 0.838/ -0.838 & IL: 0.843/ -0.843 & IL: 1.176        & IL: 1.175 \\
        \hline\hline
    \end{tabular}
      \label{table:moments_Imma}
\end{table}


\begin{table}
    \setlength{\tabcolsep}{6pt}
    \renewcommand{\arraystretch}{1.2}
    \caption{ Magnetic moments of the inequivalent Ni atoms (in $\mu_B$) for $Cmmm$-La$_{3}$Ni$_{2}$O$_{7}$-1313 for the different magnetic states depicted in Fig. 3 as a function of $U$. }
    \vspace{4pt}
    \begin{tabular}{c c c c c c c}
        \hline\hline
        $U$ (eV) & \textcolor{gray}{DSA} &
        \textcolor{magenta}{SL: AFM } &
        SL: AFM  &
        \textcolor{green}{SL: AFM} &
        \textcolor{red}{SL: AFM } &
        \textcolor{yellow}{ SL: FM} \\
                &  &
        \textcolor{magenta}{TL: AFM/0/AFM} &
         TL: AFM-C &
        \textcolor{green}{TL: AFM-G} &
        \textcolor{red}{TL: FM} &
        \textcolor{yellow}{ TL: FM} \\
        
        \hline
        \multirow{3}{*}{3}
        & SL: 0.279/ -0.279   & SL: 1.399/ -1.399 & SL: 1.403/ -1.403 & SL: 1.394/ -1.394 & SL: 1.443/ -1.443 & SL: 1.558 \\
        & OL: 1.174/ -1.174   & OL: 1.078/ -1.078 & OL: 1.075/ -1.075 & OL: 1.084/ -1.084 & OL: 1.326        & OL: 1.362 \\
        & IL: 0.523/ -0.523   & IL: 0.500/ -0.500 & IL: 0.481/ -0.481 & IL: 0.463/ -0.463 & IL: 0.904        & IL: 0.835 \\
        \hline
        \multirow{3}{*}{4}
        & SL: 0.409/ -0.409   & SL: 1.495/ -1.495 & SL: 1.481/ -1.481 & SL: 1.479/ -1.479 & SL: 1.522/ -1.522 & SL: 1.609 \\
        & OL: 1.297/ -1.297   & OL: 1.269/ -1.269 & OL: 1.217/ -1.217 & OL: 1.222/ -1.222 & OL: 1.393        & OL: 1.431 \\
        & IL: 0.655/ -0.655   & IL: 0.009/ -0.009 & IL: 0.655/ -0.655 & IL: 0.647/ -0.647 & IL: 1.077        & IL: 1.081 \\
        \hline
        \multirow{3}{*}{5}
        & SL: 0.495/ -0.495   & SL: 1.565/ -1.565 & SL: 1.552/ -1.552 & SL: 1.554/ -1.554 & SL: 1.581/ -1.581 & SL: 1.657 \\
        & OL: 1.419/ -1.419   & OL: 1.378/ -1.378 & OL: 1.336/ -1.336 & OL: 1.338/ -1.338 & OL: 1.455        & OL: 1.487 \\
        & IL: 0.699/ -0.699   & IL: 0.114/ -0.114 & IL: 0.810/ -0.810 & IL: 0.805/ -0.805 & IL: 1.156        & IL: 1.152 \\
        \hline\hline
    \end{tabular}
      \label{table:moments_Cmmm}
\end{table}

\section{\label{appendix:E} Unfolded electronic band structure of the magnetic ground state of La$_{3}$Ni$_{2}$O$_{7}$-1313}

\begin{figure}[h!]
    \centering
    \includegraphics[width=0.35\linewidth]{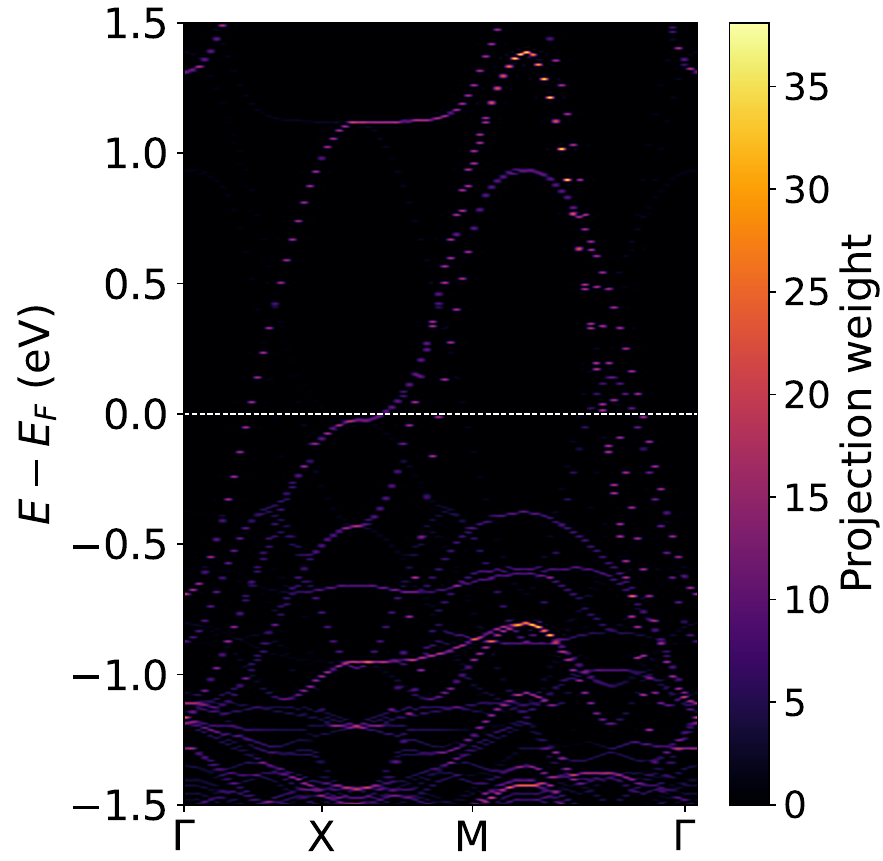}
    \caption{Unfolded electronic band structure of La$_{3}$Ni$_{2}$O$_{7}$-1313 in the magnetic ground state described in the main text (SL:AFM ; TL:FM).}
    \label{fig:unfolded_bands}
\end{figure}

Fig.~\ref{fig:unfolded_bands} shows the unfolded electronic band structure of the magnetic ground state (SL:AFM ; TL:FM) of La$_{3}$Ni$_{2}$O$_{7}$-1313 for the majority spin channel. The bands are obtained by using the KPROJ package ~\cite{chen_2018_PRB_kprojection,chen_2018_ACS_unfoldedbands} . The method implemented in KPROJ uses the projection scheme under which each $|k_{i},\epsilon_{i}>$ eigenvalue from the folded Brillouin zone  is projected to the $|K_{j},E_{j}>$ eigenvalue of the primitive Brillouin zone. The method gives the overlap in terms of weights which are plotted in the form of spectral functions.  The bands are very similar to those of La$_4$Ni$_3$O$_{10}$ \cite{li2017} apart from the reduced bandwidth due to confinement effects.

\end{document}